\documentclass[showpacs,twocolumn,prb]{revtex4-1}
\usepackage{graphicx}
\graphicspath{{figures/}} 

\newcommand{\bwt}{\begin{widetext}}
\newcommand{\ewt}{\end{widetext}}

\newcommand{\be}{\begin{equation}}
\newcommand{\ee}{\end{equation}}
\def\bea {\begin{eqnarray}}
\def\eea {\end{eqnarray}}  
\def \bk{{\bf k}}
\def \bkp{{\bf k^\prime}}

\def \iwm {i\omega_m}
\def \iwmp {i\omega_{m^\prime}}
\def \kiwm {\bk,i\omega_m}

\usepackage{hyperref}
\hypersetup{
    colorlinks=true,
    linkcolor=blue,
    filecolor=magenta,      
    urlcolor=blue,
}

\usepackage{color} 
\newcommand{\itt}{\it}
\newcommand{\black}{\textcolor{black}}

\usepackage{comment}
\def\comment#1{}
\newcommand{\com}{\black}

\usepackage{ulem}   

\begin{document}

\title{Eliashberg Theory in the Weak Coupling Limit}

\author{F. Marsiglio}
\affiliation{Department of Physics, University of Alberta, Edmonton, AB, Canada T6G~2E1}

\begin{abstract}
Eliashberg theory provides a theoretical framework for understanding the phenomenon of superconductivity
when pairing between two electrons is mediated by phonons, and retardation effects are fully accounted for.
BCS theory is often viewed as the weak coupling limit of Eliashberg theory, in spite of a handful of papers that
have pointed out that this is not so. Here we present very accurate numerical solutions in the weak coupling limit to
complement the existing analytical results, and demonstrate more convincingly the validity of this limit by extending
the analytical results to first order in the coupling constant. 
\end{abstract}

\date{\today }
\maketitle

\section{introduction}

The Eliashberg theory of superconductivity\cite{eliashberg60} provides a framework for superconductivity in which
the pairing ``glue,'' in this case phonons, is not so much a ``glue'' as a mediator of the interaction between two electrons.
In contrast, the Bardeen-Cooper-Schrieffer (BCS) theory of superconductivity\cite{bardeen57} uses a pairing potential to model the
attractive interaction between two electrons. Being a potential the interaction is instantaneous, although retardation effects
are mimicked through a cutoff in the potential, albeit in wave-vector space and not in frequency space. 

Eliashberg theory is sometimes referred to as the ``strong-coupling'' extension of BCS theory. The reason no doubt is that
superconducting materials in which retardation effects play a significant role (e.g. Pb and Hg) also tend to have a stronger
electron-phonon coupling than those in which their role is minor (e.g. Al). Furthermore, in Eliashberg theory the quasiparticles
have a finite width and their residue is no longer unity, and both of these factors contributed to this misnomer. 
In fact, both Eliashberg and BCS theory are weak coupling theories in the sense that the starting point is a Fermi sea of
electrons, so what really delineates the two is that the former explicitly includes retardation effects while the latter does
not. Formally, the strong coupling limit in both these theories can be investigated (and have been --- see Refs.~[\onlinecite{thouless60,swihart62}] for BCS and Refs.~[\onlinecite{allen75,marsiglio91}] for Eliashberg theory). However,
particularly at finite temperature these calculations are beyond the limit of validity of the formulation, as the condensation of
preformed pairs, whose constituents {\it do not} form a Fermi sea, is the physically relevant process, which is not described by
these theoretical frameworks.\cite{remark_leggett}

There is a tacit understanding that the weak coupling limit of both theories converge to the same limits. This belief has
been reinforced, for example, in studies of universal BCS constants like the gap ratio\cite{mitrovic84} and the normalized
specific heat jump.\cite{marsiglio86} In these and other cases\cite{carbotte90} universal BCS constant show deviations within Eliashberg
theory that eventually achieve the BCS value as the coupling becomes weaker.

That this is {\it not universally} the
case was first noted by Karakozov et al.\cite{karakozov76} In fact they showed that a correction to the BCS pre-factor appears
in the weak coupling limit of Eliashberg theory for the determination of $T_c$, the superconducting critical temperature itself. This is
an important observation and merits further investigation. In this paper we will re-derive this result for $T_c$ (on the imaginary
axis following Ref.~[\onlinecite{wang13}]) and we will also derive an improved analytical form for the order parameter as well. 
Remarkably the order parameter is not at
all a constant over a frequency range of the typical phonon frequency, as modelled both in BCS theory, and even in Eliashberg theory with
the so-called square-well model for the electron-phonon interaction introduced by McMillan.\cite{mcmillan68}

Note that in this study we examine corrections to BCS that arise entirely within Eliashberg theory; there are 
a number of additional contributions that have an effect on the pre-factor, 
for example, that of 
Kohn and Luttinger,\cite{kohn65,wang13b} but we do not address those here.

We proceed as follows. First we provide a quick synopsis of Eliashberg theory. We take some effort to review the so-called
``standard'' approximations to arrive at the self-consistent equations for the order parameter as a function of Matsubara frequency
only. As emphasized in Ref.~[\onlinecite{allen82}] these approximations are quite controlled precisely in the weak coupling limit,
and have properly been avoided or modified for further more recent refinements in the theory.\cite{chubukov08} Here, however,
these approximations rest on solid ground. We then present both numerical and analytical solutions to the gap function, first
following Wang and Chubukov\cite{wang13} in the case where renormalization effects are neglected, and then in the case where they are accounted
for. While $T_c$ is unaffected (except for the usual mass renormalization term, $1+\lambda$), the high frequency dependence
of the gap function to first order in $\lambda$ is indeed changed, as described in more detail below. 
We conclude with a summary in the final Section.

\section{Eliashberg Theory Formalism}

 The Eliashberg equations are\cite{marsiglio08}
\bwt

\bea
Z({\bf k},i\omega_m) = 1 + {1 \over N\beta}
\sum_{\bkp,m^\prime}
{\lambda_{\bk \bkp}(\iwm - \iwmp) \over g_{\epsilon_F}}
{ \bigl(\omega_{m^\prime} / \omega_m \bigr)
Z({\bf k^\prime},i\omega_{m^\prime}) \over 
\omega_{m^\prime}^2 Z^2({\bf k^\prime},i\omega_{m^\prime}) + 
\bigl( \epsilon_{\bf k^\prime} - \mu + \chi({\bf k^\prime},i\omega_{m^\prime})
\bigr)^2  + \phi^2({\bf k^\prime},i\omega_{m^\prime})}
\label{ga1}
\\                                                          
\chi({\bf k},i\omega_m) = - {1 \over N\beta}
\sum_{\bkp,m^\prime}
{\lambda_{\bk \bkp}(\iwm - \iwmp)  \over g_{\epsilon_F}} 
{ \epsilon_{\bf k^\prime} - \mu + \chi({\bf k^\prime},i\omega_{m^\prime}) 
\over
\omega_{m^\prime}^2 Z^2({\bf k^\prime},i\omega_{m^\prime}) +
\bigl( \epsilon_{\bf k^\prime} - \mu + \chi({\bf k^\prime},i\omega_{m^\prime})
\bigr)^2  + \phi^2({\bf k^\prime},i\omega_{m^\prime})}
\label{ga2}   
\eea
along with the equation for the order parameter: 
\be
\phi({\bf k},i\omega_m) = {1 \over N\beta}
\sum_{\bkp,m^\prime}
{\lambda_{\bk \bkp}(\iwm - \iwmp)  \over g_{\epsilon_F}} 
{ \phi({\bf k^\prime},i\omega_{m^\prime})
\over
\omega_{m^\prime}^2 Z^2({\bf k^\prime},i\omega_{m^\prime}) +
\bigl( \epsilon_{\bf k^\prime} - \mu + \chi({\bf k^\prime},i\omega_{m^\prime})
\bigr)^2 + \phi^2({\bf k^\prime},i\omega_{m^\prime})}.
\label{ga3}     
\ee
These are supplemented with the electron number equation, which determines
the chemical potential, $\mu$:
\bea
\rho & = & 1 - {2 \over N\beta} \sum_{\bkp,m^\prime} 
{ \epsilon_{\bf k^\prime} - \mu + \chi({\bf k^\prime},i\omega_{m^\prime})
\over
\omega_{m^\prime}^2 Z^2({\bf k^\prime},i\omega_{m^\prime}) +
\bigl( \epsilon_{\bf k^\prime} - \mu + \chi(({\bf k^\prime},i\omega_{m^\prime})
\bigr)^2  + \phi^2({\bf k^\prime},i\omega_{m^\prime})}.
\label{ga4}
\eea

\ewt
Here, $N$ is the number of lattice sites, $\beta \equiv 1/(k_BT)$, where $k_B$ is the Boltzmann constant and $T$ is the temperature,
$\mu$ is the chemical potential, and $g_{\epsilon_F}$ is the electronic density of states at the Fermi level
in the band. 
The energy $\epsilon_{\bf k}$ is the electronic dispersion of this band (a single band is assumed for simplicity). The equations are written on
the imaginary frequency axis, and are functions of the Fermion Matsubara frequencies, $\omega_m \equiv \pi k_B T (2m-1)$, with $m$
an integer. Similarly the Boson Matsubara frequencies are given by $\nu_n \equiv 2 \pi k_B T n$, where $n$ is an integer
The functions $Z({\bf k},i\omega_m))$ and $\chi({\bf k},i\omega_m)$ are related to the electron self energy through
\cite{allen82}
\begin{eqnarray}
i\omega_m \bigl[ 1 - Z({\bf k},i\omega_m) \bigr]  & \equiv &
{1 \over 2} \bigl[ \Sigma(\kiwm) - \Sigma({\bf k},-i\omega_m) \bigr]
\nonumber \\
\chi(\kiwm) & \equiv & 
{1 \over 2} \bigl[ \Sigma(\kiwm) + \Sigma({\bf k},-i\omega_m) \bigr]
\label{even_odd}
\end{eqnarray}
where $Z$ and $\chi$ are both even functions of $i\omega_m$ (and, as
we've assumed from the beginning, ${\bf k}$). The function $\phi({\bf k},i\omega_m))$ is the so-called pairing function, and
is related to the electronic anomalous Green function. These equations relate these three functions to one another through
the electron-phonon propagator, contained in
\be
\lambda_{\bk \bkp} (z) \equiv \int_0^\infty {2 \nu \alpha_{\bk \bkp}^2
F(\nu) \over \nu^2 - z^2} d \nu
\label{lambda_z}
\ee
\noindent with $\alpha_{\bk \bkp}^2 F(\nu)$ the so-called Eliashberg function. In what follows we will assume that
the phonon spectrum is given by an Einstein spectrum and that the coupling is wave vector independent. Therefore,
\be
\alpha_{\bk \bkp}^2F(\nu) = (\lambda \omega_E/2) \delta(\nu - \omega_E)
\label{a2f_ein}
\ee
and the kernel, Eq.~(\ref{lambda_z}), is written as
\be
\lambda(i\nu_n)  = {\lambda \omega_E^2 \over \omega_E^2 + \nu_n^2}
\label{lambda_n}
\ee
where the constant $\lambda$ is the dimensionless electron-phonon coupling constant and $\omega_E$ is the Einstein
phonon frequency. Normally a direct Coulomb repulsion is also included in the pairing equation; we omit this here since we
want to focus on the effects of retardation. 
The fourth equation, Eq.~(\ref{ga4}), is used to determine the chemical potential given an electron
density $\rho$, but in this work we will assume particle-hole symmetry; then $\mu = 0$ always and this equation 
is not used, with $\rho$ no longer relevant. Similarly, $\chi({\bf k},i\omega_m)$ is identically zero.
We furthermore assume that the electronic density of states 
is essentially a constant over the energy range of interest, and set it equal to the value of
the density of states at the Fermi level, $g(\mu) \approx g_{\epsilon_F}$.  With these assumptions the equations simplify considerably, and none of
the functions has any wave vector dependence, i.e. they are solely functions of Matsubara frequency, $\omega_m$.
Focussing our attention on the onset of superconductivity and the critical temperature, we linearize the equations and obtain
\be
Z(i\omega_m)  =  1+ {\pi T_c \over \omega_m} \sum_{m^\prime} \lambda(i\omega_m - i\omega_{m^\prime})
{\rm sgn}(\omega_{m^\prime}).
\label{gb1}
\ee
\be
\phi(i\omega_m)  =  \pi T_c \sum_{m^\prime} 
\lambda(i\omega_m - i\omega_{m^\prime})
{\phi(i\omega_{m^\prime}) \over |\omega_{m^\prime}| Z(i\omega_{m^\prime})}.
\label{gb3}
\ee
The case of a constant density of states but with a finite bandwidth was
examined in Ref.~[\onlinecite{marsiglio92}]; it is apparent from that work that in the weak coupling limit this
bandwidth is irrelevant for $T_c$.\cite{remark_other} 
Equations~(\ref{gb1}) and (\ref{gb3}) are the ``standard'' linearized Eliashberg equations, valid for infinite electronic bandwidth.
The function $Z(i\omega_m)$ can be determined in closed form; we obtain, for $\omega_m > 0$ (since both $Z$ and $\phi$
are even real functions of $\omega_m$),
\be
Z(i\omega_m) = 1 + {\pi k_B T_c \over \omega_m} \left\{ \lambda + 2 \sum_{n=1}^{m-1} \lambda(i\nu_n) \right\}.
\label{zz_explicit}
\ee
It is also standard practice to define a ``gap function,'' $\Delta(i\omega_m) \equiv \phi(i\omega_m)/Z(i\omega_m)$,
so that the remaining equation to determine $T_c$ is
\be
Z(i\omega_m) \Delta(i\omega_m)  =  \pi T_c \sum_{m^\prime = -\infty}^{+\infty} 
\lambda(i\omega_m - i\omega_{m^\prime})
{\Delta(i\omega_{m^\prime}) \over |\omega_{m^\prime}|}.
\label{gap}
\ee
Equations~(\ref{zz_explicit}) and (\ref{gap}) were first solved in this form in Refs.~(\onlinecite{owen71,bergmann73,rainer74}),
and have been solved many times since. 

As mentioned in the Introduction, one can examine Eliashberg theory in limiting cases of weak coupling ($\lambda \rightarrow 0$)
and strong coupling $\lambda \rightarrow \infty$. Interestingly, Eq.~(\ref{gap}) is readily solved numerically in the latter
limit (see e.g. Refs.~(\onlinecite{allen75,carbotte86,marsiglio91})), but not so easily in the former limit. 
Approximate forms like the square-well model were first used by McMillan,\cite{mcmillan68} and adopted in subsequent reviews.\cite{allen82,marsiglio08}
In the end however, McMillan and others adopted phenomenological pre-factors, whose justification is now more readily understood after
Karakozov et al.\cite{karakozov76b} solved the gap equation on the real axis with an iterative method
and obtained the result that $T_c$ attains a pre-factor significantly different than that obtained with BCS theory.\cite{remark_prefactor}
We will first re-derive this result on the imaginary axis\cite{wang13} and determine an analytical approximation for the 
gap function.

The equation for $T_c$ within BCS theory is (we now set $k_B = 1$ and $\hbar = 1$)
\be
T_c = 1.13 \omega_E \exp{(-1/\lambda)}
\label{bcs_tc}
\ee
where $\lambda \equiv g_{\epsilon_F} |V|$, with $|V|$ some attractive and instantaneous potential between two electrons.
The inclusion of the renormalization, $Z$, modifies this equation to read
\be
T_c = 1.13 \omega_E \exp{(-(1+ \lambda)/\lambda)}.
\label{bcs_tc_zz}
\ee
One can immediately write this like Eq.~(\ref{bcs_tc}) but with reduced pre-factor $1.13e^{-1}$. This is {\it not} what
is meant when we stated that the pre-factor in Eliashberg theory is actually modified from the BCS result --- but rather
an additional change occurs.

\section{Un-renormalized Eliashberg Theory}

\subsection{Improved $T_c$ in the $\lambda \rightarrow 0$ limit}

To emphasize this latter point we first examine the Eliashberg $T_c$ equation, Eq.~(\ref{gap}) with $Z(i\omega_m) \equiv 1$,
i.e.
\be
\Delta(i\omega_m)  =  \pi T_c \sum_{m^\prime = -\infty}^{+\infty} 
\lambda(i\omega_m - i\omega_{m^\prime})
{\Delta(i\omega_{m^\prime}) \over |\omega_{m^\prime}|}.
\label{gap_nozz}
\ee
We immediately caution that this is a dangerous step to make, as emphasized by Cappelluti and Ummarino.\cite{cappelluti07}
In fact this choice results in unstable equations for $\lambda > 1$. Since we are interested only in the weak coupling limit $\lambda << 1$,
Eq.~(\ref{gap_nozz}) remains stable.
In what follows we make use of the fact that even within Eliashberg theory the structure of Eq.~(\ref{bcs_tc}) remains intact,
so that $T_c/\omega_E \approx e^{-1/\lambda} << 1$ for the weak coupling case. 
The impact on $\Delta(\omega_m)$ is, however, a little more subtle and a discussion of this case will be deferred to the next section.

For now, with $Z(\omega_m) = 1$, we begin by writing Eq.~(\ref{gap_nozz}) as
\bwt

\bea
\Delta(i\omega_m)  &=&  \lambda \pi \bar{T}_c \sum_{m^\prime = -\infty}^{+\infty} 
{1 \over 1 + (\bar{\omega}_m - \bar{\omega}_{m^\prime})^2}
{\Delta(i\omega_{m^\prime}) \over |\bar{\omega}_{m^\prime}|} \label{gap1a}\\
&=&  {1 \over 1 + {\bar{\omega}_m}^2} \lambda \pi \bar{T}_c \sum_{m^\prime = -\infty}^{+\infty} 
\left\{ 1 + {2 \bar{\omega}_m \bar{\omega}_{m^\prime} - {\bar{\omega}_{m^\prime}}^2 
\over 1 + (\bar{\omega}_m - \bar{\omega}_{m^\prime})^2} \right\}
{\Delta(i\omega_{m^\prime}) \over |\bar{\omega}_{m^\prime}|}
\label{gap1b}
\eea

\ewt
where $\bar{Q} \equiv Q/\omega_E$, and in the second line we have added and subtracted the factor $1/(1 + {\bar{\omega}_m}^2)$.
Eq.~(\ref{gap1b}) makes it clear that one can write
\be
\Delta(i\omega_m)  =  {1 \over 1 + {\bar{\omega}_m}^2} \left( 1 + \lambda f(\omega_m) \right).
\label{gap2}
\ee
This equation looks like a perturbative expansion in $\lambda$; if we neglect $f(\omega_m)$, and {\it further}
neglect the second complicated-looking term in Eq.~(\ref{gap1b}), we obtain simply
\be
1 \approx \lambda \pi \bar{T}_c \sum_{m^\prime = -\infty}^{+\infty} {1 \over |\bar{\omega}_{m^\prime}|}
{1 \over 1 + {\bar{\omega}_{m^\prime}}^2}  \equiv \lambda I_0,
\label{gap3}
\ee
where $I_0$ can be evaluated in terms of the asymptotic expansion of digamma functions\cite{abramowitz72,olver10}, as
\be
I_0 \approx {\rm ln}\left({1.13 \omega_E \over T_c}\right) - {\pi^2 \over 6} \left( {T_c \over \omega_E}\right)^2.
\label{iosum}
\ee
Upon neglecting the last term, the result is that we obtain the usual BCS $T_c$ equation given by Eq.~(\ref{bcs_tc}). In fact
it is inconsistent to neglect the complicated-looking second term in Eq.~(\ref{gap1b}). Thus, while still neglecting the corrections
proportional to $f(\omega_m)$, a more accurate version of Eq.~(\ref{gap3}) more correctly contains an additional term, 
so this equation reads
\be
1 \approx \lambda I_0 + \lambda \pi \bar{T}_c \sum_{m^\prime = -\infty}^{+\infty} 
{1 \over 1 + {\bar{\omega}_{m^\prime}}^2}
{2 \bar{\omega}_m {\rm sgn}(\bar{\omega}_{m^\prime}) - |{\bar{\omega}_{m^\prime}}| 
\over 1 + (\bar{\omega}_m - \bar{\omega}_{m^\prime})^2}.
\label{gap4}
\ee
This equation is clearly an approximation since the second term has a dependence on $\omega_m$; this reflects the approximation
inherent in Eq.~(\ref{gap2}) when $f(\omega_m)$ is neglected. Nonetheless, we multiply both sides of Eq.~(\ref{gap4})
by $\pi \bar{T}_c \left\{1/|\bar{\omega}_m|\right\} \left\{1/(1 + \bar{\omega}_m^2)\right\}$
and sum over all values of $m$, to obtain
\bwt

\be
I_0 = \lambda I_0^2 - \lambda (\pi \bar{T}_c)^2 \sum_{m,m^\prime = -\infty}^{+\infty} {1 \over 1 + {\bar{\omega}_{m^\prime}}^2}
{1 \over |\bar{\omega}_m|} {1 \over 1 + \bar{\omega}_m^2}
\left\{{|{\bar{\omega}_{m^\prime}}| - 2 \bar{\omega}_m {\rm sgn}(\bar{\omega}_{m^\prime}) 
\over 1 + (\bar{\omega}_m - \bar{\omega}_{m^\prime})^2}\right\}.
\label{gap5a}
\ee
Use\cite{wang13}
\be
{1 \over 1 + (\bar{\omega}_m - \bar{\omega}_{m^\prime})^2} = {1 \over 1 + \bar{\omega}_{m^\prime}^2}
+ \left\{{1 \over 1 + (\bar{\omega}_m - \bar{\omega}_{m^\prime})^2} - {1 \over 1 + \bar{\omega}_{m^\prime}^2}\right\}.
\label{gap5b}
\ee

\ewt
to replace the term in braces in Eq.~(\ref{gap5a}). The first term (proportional to $|\bar{\omega}_{m^\prime}|$ in the numerator
of the sum in this equation is seen to contain a singular part as $T_c \rightarrow 0$ (since a denominator proportional to 
$|\bar{\omega}_m|$ remains), which in effect
offsets the diminution of the $\lambda$ in the pre-factor. The singular part is extracted by adding and subtracting
$\left\{1/(1 + \bar{\omega}_{m^\prime}^2)\right\}$ as indicated in Eq.~(\ref{gap5b}). Then the first term
contains the singular part, while the remainder is of order unity, and therefore remains small due to the $\lambda$ pre-factor.
Eq.~(\ref{gap5a}) then becomes
\be
I_0 = \lambda I_0^2 - I_0/2,
\label{tcapp1}
\ee
where we have used the fact that
\be
I_4 \equiv (\pi \bar{T}_c) \sum_{m = -\infty}^{+\infty} 
{|\bar{\omega}_m| \over (1 + \bar{\omega}_m^2)^2} \approx {1 \over 2}.
\label{summ_i4}
\ee
Following Refs.~[\onlinecite{karakozov76,wang13}] we solve Eq.~(\ref{tcapp1}) to obtain
\be
T_c = {1.13 \over \sqrt{e}} \omega_E \exp{(-1/\lambda)}
\label{bcs_tc_app1}
\ee
in contrast to Eq.~(\ref{bcs_tc}). 

\begin{figure}[tp]
\begin{center}
\includegraphics[height=2.8in,width=2.8in,angle=-90]{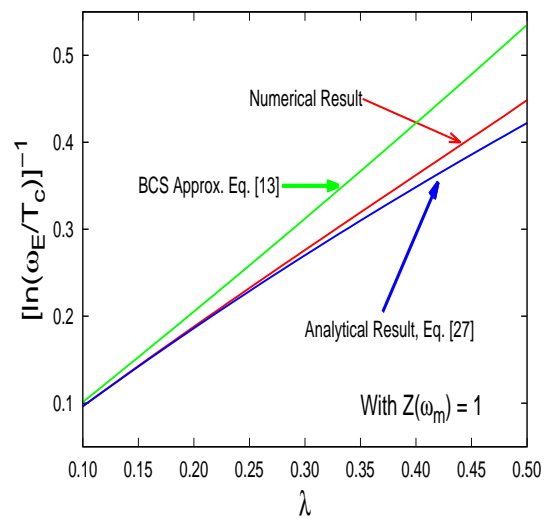}
\end{center}
\caption{A plot of $[{\rm ln}(\omega_E/T_c)]^{-1}$ vs. $\lambda$. Numerical results are shown in red; the usual BCS approximation,
Eq.~(\ref{bcs_tc}), is given by the green curve, while the improved estimate given by Eq.~(\ref{bcs_tc_app1})  is shown in blue.
This latter result becomes essentially exact for $\lambda {{ \atop <} \atop {\approx \atop }} 0.2$.
}
\label{fig1_chubukov}
\end{figure}

Fig.~1 shows results from un-renormalized Eliashberg theory
(solved numerically), along with the BCS result from Eq.~(\ref{bcs_tc}) and the improved result
from Eq.~(\ref{bcs_tc_app1}). In particular we plot $[{\rm ln}(\omega_E/T_c)]^{-1}$ vs. $\lambda$.
The numerical results are given as a red curve as indicated, while both the BCS approximation Eq.~(\ref{bcs_tc}) and the
improved result from Eq.~(\ref{bcs_tc_app1}) are given by green and blue curves, respectively, as indicated. 
It is clear that the improved result is essentially exact for
the weakest electron-phonon couplings shown. 

\subsection{Improved gap function in the $\lambda \rightarrow 0$ limit}

\begin{figure}[tp]
\begin{center}
\includegraphics[height=2.8in,width=2.8in,angle=-90]{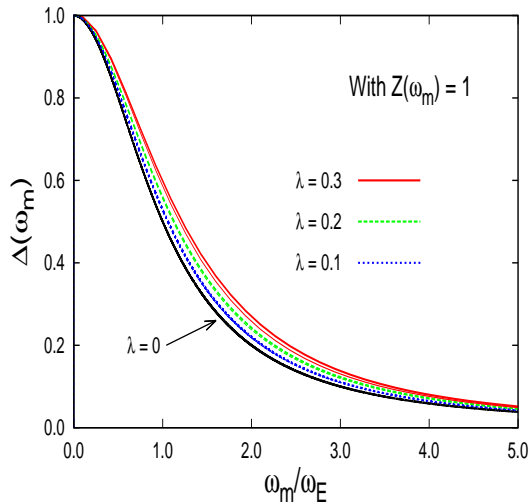}
\end{center}
\caption{A plot of $\Delta(\omega_m)$ vs. $\bar{\omega}_m \equiv \omega_m/\omega_E$ for $\lambda = 0.3, 0.2$, and $0.1$
as indicated, obtained numerically. Also shown is the approximation given by Eq.~(\ref{gap2}) 
with $f(\omega_m) = 0$. It is clear that deviations from this limiting result certainly
exist, but the numerical results are certainly trending towards this weak coupling result. Note that all curves shown are actually a discrete
set of points, determined at the Matsubara frequencies, but curves have been drawn for better presentation. In reality only the results
for $\lambda = 0.3$ are readily discerned as a discrete set. For reference, the two-square well model would be a step function
with value of unity for $0 < \bar{\omega}_m < 1$ and value zero beyond. The numerical values of $T_c/\omega_E$ for each 
of these cases is $T_c/\omega_E = 0.026744 (\lambda = 0.3)$, 
$0.004900 (\lambda = 0.2)$, and 
$0.000032 (\lambda = 0.1)$. Note that an improved approximation to first order in $\lambda$, given by Eq.~(\ref{gap8}) with $g_1(\omega_m)$
provided by Eq.~(\ref{fwm}), is shown with a thin curve of the same colour for each value of $\lambda$. The result is discernible from the
numerical result only in the case of $\lambda = 0.3$. }
\label{fig2_chubukov}
\end{figure}

One of the physical features of the square well model referred to in the previous section is that the
gap function is a constant for a range of energies equal to the phonon frequency (here, $\omega_E$)
to either side of the Fermi energy. This is already not true with the approximation provided by Eq.~(\ref{gap2}),
even with the neglect of $f(\omega_m)$. In Fig.~2 we show with thick curves the numerical result for the gap function for several
weak values of the coupling parameter, $\lambda$, along with the result from Eq.~(\ref{gap2}) with $f(\omega_m) \equiv 0$. 
This latter result,
with $f(\omega_m) = 0$, is independent of $\lambda$ and will presumably be correct in the strict $\lambda \rightarrow 0$
limit. Fig.~2 clearly confirms that the numerical results are indeed trending towards this result.

In an effort to further improve this result and refine our understanding of the weak coupling limit, we proceed to determine
$f(\omega_m)$ at least as a correction to zeroth order in $\lambda$ (and thus an overall correction to the gap function to 
first order in $\lambda$). For this purpose we substitute Eq.~(\ref{gap2}) into Eq.~(\ref{gap1b}); upon isolating $f(\omega_m)$
we obtain
\be
f(\omega_m) = c  - g_1(\omega_m) - \lambda g_2(\omega_m)
\label{gap6}
\ee
where $c$ is a constant given by
\be
c = -{1 \over \lambda} + I_0 + \lambda \pi \bar{T}_c \sum_{m^\prime = -\infty}^{+\infty} {f(\omega_{m^\prime})
\over |{\bar{\omega}_{m^\prime}}|} {1 \over 1 + {\bar{\omega}_{m^\prime}}^2}  
\label{gap7a}
\ee
and
\be
g_1(\omega_m) = \pi \bar{T}_c \sum_{m^\prime = -\infty}^{+\infty} {1 \over 1 + {\bar{\omega}_{m^\prime}}^2}
\left\{{|{\bar{\omega}_{m^\prime}}| - 2 \bar{\omega}_m {\rm sgn}(\bar{\omega}_{m^\prime}) 
\over 1 + (\bar{\omega}_m - \bar{\omega}_{m^\prime})^2}\right\}
\label{gap7b}
\ee
and
\be
g_2(\omega_m) = \pi \bar{T}_c \sum_{m^\prime = -\infty}^{+\infty} {f(\omega_{m^\prime}) \over 1 + {\bar{\omega}_{m^\prime}}^2}
\left\{{|{\bar{\omega}_{m^\prime}}| - 2 \bar{\omega}_m {\rm sgn}(\bar{\omega}_{m^\prime}) 
\over 1 + (\bar{\omega}_m - \bar{\omega}_{m^\prime})^2}\right\}
\label{gap7c}
\ee
are two functions of $\omega_m$.
Both $g_1(\omega_m)$ and $g_2(\omega_m)$ are non-singular as $\lambda \rightarrow 0$. By this we mean that a
$1/|\bar{\omega}_{m^\prime}|$ term is absent (as opposed to $I_0$, for example, the sum multiplying 
$\lambda$ in Eq.~(\ref{gap3})); this means both of these functions are of order unity. Since $\lambda$ premultiplies
$g_2(\omega_m)$, $g_2$ can be ignored, bearing in mind we wish to retain terms in $f(\omega_m)$ of order unity or
better. The resulting expression for the constant $c$ is
\be
c = -{1 \over \lambda} + I_0 + \lambda c I_0 - \lambda \left( {1 \over 2} I_0 + c^\prime \right)
\label{c_eqn}
\ee
where $c^\prime$ is a constant obtained numerically from the sum in Eq.~(\ref{gap7a}) with $g_1(\omega_m)$
substituted as part of $f(\omega_m)$.  In any event $c^\prime$ is irrelevant as it is multiplied by $\lambda$
and enters only at higher order in $\lambda$. The result is $c = 1/2$, obtained already through the eigenvalue
equation, Eq.~(\ref{tcapp1}). This results in an improved $T_c$ result given by Eq.~(\ref{bcs_tc_app1}).

This leaves the explicit expression for $g_1(\omega_m)$ in Eq.~(\ref{gap7b}); this can be evaluated to order $(T_c/\omega_E)^2$
through the properties of digamma functions,\cite{abramowitz72,olver10}
\be
g_1(\omega_m) = {1 \over 4 + \bar{\omega}_m^2}\left\{  
{2 - \bar{\omega}_m^2 \over \bar{\omega}_m}{\rm tan}^{-1}\bar{\omega}_m  - {3 \over 2} {\rm ln}(1 + \bar{\omega}_m^2) \right\},
\label{fwm}
\ee
and we now have a more accurate explicit expression for the gap function,
\be
\Delta(\omega_m) = {1 \over 1 + \bar{\omega}_m^2} \left( 1 + \lambda ( {1 \over 2} - g_1(\omega_m) ) \right),
\label{gap8}
\ee
valid to order $\lambda$. Three thin curves showing this result for $\lambda = 0.1, 0.2$ and $0.3$ on the scale of 
Fig.~2 are essentially indistinguishable from the numerical
results, and show that up to $\lambda \approx 0.3$ at least, Eq.~(\ref{gap8}), with $g_1(\omega_m)$ from Eq.~(\ref{fwm}), is
very accurate for small but non-zero values of $\lambda$.

To better appreciate the remaining discrepancies, we show in Fig.~3
results for the deviation from the universal result,
\be
{\Delta}_0(\omega_m) = {1\over {1+ \bar{\omega}_m^2}},
\label{delta0}
\ee
defined as $\delta \Delta_{\rm num}(\omega_m) \equiv {\Delta}_{\rm num}(\omega_m) - {\Delta}_0(\omega_m)$, where ${\Delta}_{\rm num}(\omega_m)$ 
refers to the numerical solution\cite{remark_normalization} and $\delta \Delta_{\rm ana}(\omega_m) \equiv {\Delta}_{\rm ana}(\omega_m) - {\Delta}_0(\omega_m)$, where 
${\Delta}_{\rm ana}(\omega_m)$ refers to the analytical solution given by Eq.~(\ref{gap8}). 
The remaining discrepancies for the gap function are of order $\lambda^2$. 
At this point we return to the theory with $Z(\omega_m) \ne 1$ and indicate the places where the description differs from
the one just provided.

\section{Eliashberg Theory with renormalization}

In this section we provide solutions for Eq.~(\ref{gap}), with account of Eq.~(\ref{zz_explicit}). The numerical procedure
is fairly straightforward, and follows what we did earlier. A noteworthy nuance is that the $m=m^\prime$ term on the right side
of Eq.~(\ref{gap}) no longer contributes --- it is precisely cancelled by a term on the left that arises upon 
substituting Eq.~(\ref{zz_explicit}) into Eq.~(\ref{gap}), and this is a manifestation of the lack of effect of impurities on
superconducting $T_c$, a fact pointed out by Anderson in Ref.~[\onlinecite{anderson59}]. In any event this
is properly accounted for in both the numerical and analytical results, and manifests itself not just in $T_c$, but
also in the actual functional dependence of the gap function, as we shall see below.

\begin{figure}[tp]
\begin{center}
\includegraphics[height=2.8in,width=2.8in,angle=-90]{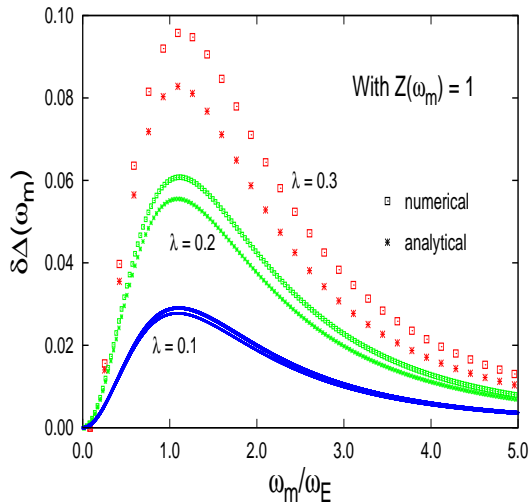}
\end{center}
\caption{A plot of the deviation from ${\Delta}_0(\omega_m)$ [see Eq.~(\ref{delta0})] 
given by the numerical results (shown with squares) and by the
analytical results (shown with asterisks, for the three different values of $\lambda$ as indicated in the figure and through the colour
scheme. In all cases the first order correction to the gap function obtained analytically through Eq.~(\ref{fwm}) very accurately accounts
for the discrepancy from ${\Delta}_0(\omega_m)$, which was not discernible in the previous figure. Note that for the 2 lowest
values of $\lambda$ only a subset of the Matsubara frequencies was used in the figure; otherwise the results would have appeared
as a continuous curve.}
\label{fig3_chubukov}
\end{figure}
\begin{figure}[tp]
\begin{center}
\includegraphics[height=2.8in,width=2.8in,angle=-90]{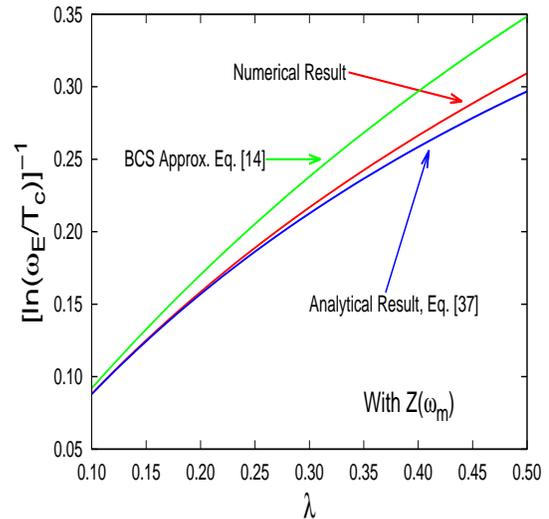}
\end{center}
\caption{A plot of $[{\rm ln}(\omega_E/T_c)]^{-1}$ vs. $\lambda$ for the case where the normal state renormalization provided
by $Z(\omega_m)$ is accounted for. Numerical results are shown in red; the usual BCS approximation,
Eq.~(\ref{bcs_tc_zz}), is given by the green curve, while the improved estimate given by Eq.~(\ref{bcs_tc_zz_app1})  is shown in blue.
This latter result becomes essentially exact for $\lambda {{ \atop <} \atop {\approx \atop }} 0.2$, and the improvement is similar to
that obtained in Fig.~1.
}
\label{fig4_chubukov}
\end{figure}
\begin{figure}[tp]
\begin{center}
\includegraphics[height=2.8in,width=2.8in,angle=-90]{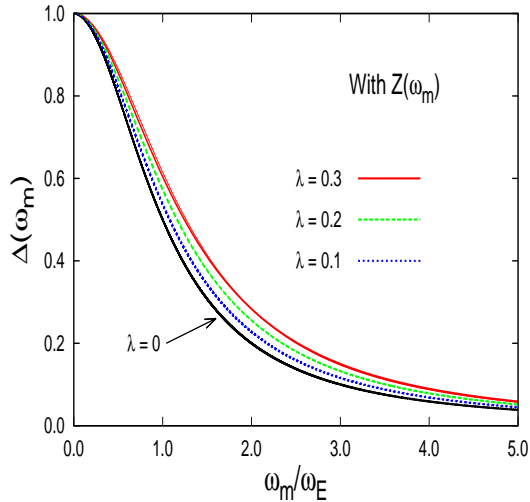}
\end{center}
\caption{Similar to Fig.~2, a plot of $\Delta(\omega_m)$ vs. $\bar{\omega}_m \equiv \omega_m/\omega_E$ for $\lambda = 0.3, 0.2$, and $0.1$ as indicated, obtained numerically (thick curves), now with the full expression for $Z(\omega_m)$ included. 
Also shown is the $\lambda \rightarrow 0$ approximation given by $1/(1 + \bar{\omega}_m^2)$ as in Fig.~2. As in that
case, deviations from this limiting result are apparent, but the numerical results are certainly trending towards this 
weak coupling result. Note that all curves shown are actually a discrete
set of points, determined at the Matsubara frequencies, but continuous 
curves have been drawn for better presentation. In reality only the results
for $\lambda = 0.3$ are readily discerned as a discrete set. In this case also, the two-square well model would be a step function
with value of unity for $0 < \bar{\omega}_m < 1$ and value zero beyond. The numerical values of $T_c/\omega_E$ for each 
of these cases is $T_c/\omega_E = 0.009923 (\lambda = 0.3)$, 
$0.001821 (\lambda = 0.2)$, and 
$0.000012 (\lambda = 0.1)$. Note that an improved approximation to first order in $\lambda$, given by Eq.~(\ref{gapzz3}), is 
shown with a thin curve of the same colour for each value of $\lambda$. The result is again barely discernible from the
numerical result only in the case of $\lambda = 0.3$.
}
\label{fig5_chubukov}
\end{figure}
\begin{figure}[tp]
\begin{center}
\includegraphics[height=2.8in,width=2.8in,angle=-90]{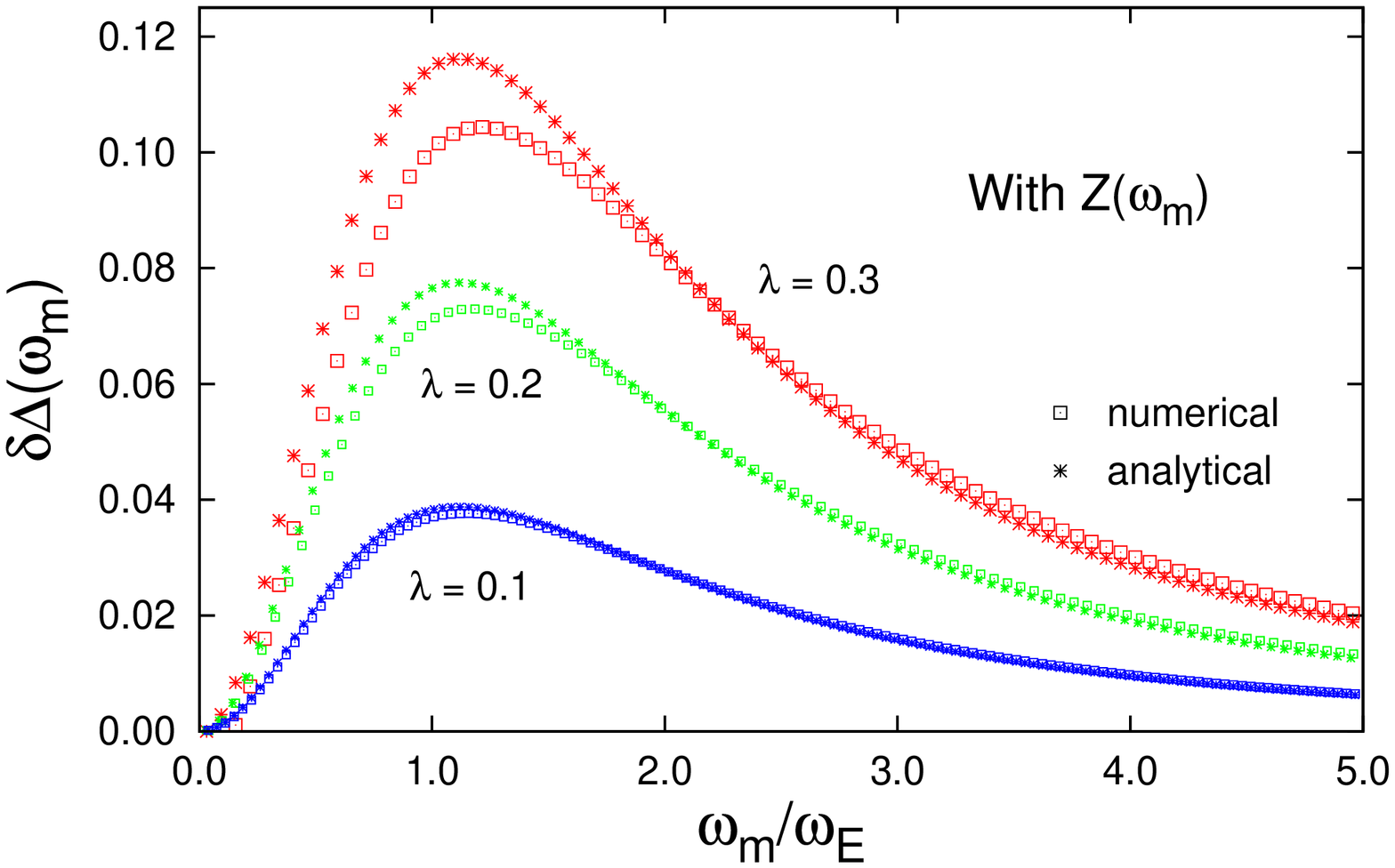}
\end{center}
\caption{Similar to Fig.~3, a plot of the deviation from ${\Delta}_0(\omega_m)$ [see Eq.~(\ref{delta0})] 
given by the numerical results (shown with squares) and by the
analytical results (shown with asterisks, for the three different values of $\lambda$ as indicated in the figure and through the colour
scheme. In all cases the first order correction to the gap function obtained analytically through Eq.~(\ref{fwm}) very accurately accounts
for the discrepancy from ${\Delta}_0(\omega_m)$; this discrepancy was not so discernible in Fig.~5. Note that for the 2 lowest
values of $\lambda$ only a subset of the Matsubara frequencies was used in the figure; otherwise the results would have appeared
as a continuous curve.}
\label{fig6_chubukov}
\end{figure}

The difference with the previous section is that $Z(\omega_m)$ is now included. The sum in Eq.~(\ref{gb1}) is readily evaluated
in terms of digamma functions.\cite{abramowitz72,olver10} Omitting terms of order $T_c/\omega_E$,
we readily obtain
\begin{equation}
Z(\omega_m) \approx 1 + \lambda {1 \over \bar{\omega}_m} {\rm tan}^{-1}\bar{\omega}_m,
\label{zz_app}
\end{equation}
which interpolates smoothly from $(1+\lambda)$ at low frequencies to unity at high frequencies.
Including this in the steps leading to Eq.~(\ref{gap2}) we obtain here instead
\begin{equation}
\Delta(\omega_m)  =  {1 \over 1 + {\bar{\omega}_m}^2} \left( 1 + \lambda \left[ f_Z(\omega_m)  - {1 \over |\bar{\omega}_m|}  
{\rm tan}^{-1}|\bar{\omega}_m| \right] \right).
\label{gapzz2}
\end{equation}
\bigskip
Following with the same type of analysis as that leading to Eq.~(\ref{bcs_tc_app1}) and to Eq.~(\ref{gap8}) we find here
that
\be
T_c = {1.13 \over \sqrt{e}} \omega_E \exp{(-(1+ \lambda)/\lambda)}.
\label{bcs_tc_zz_app1}
\ee
and
\be
f_Z(\omega_m) = {3 \over 2} - g_1(\omega_m),
\label{fzz}
\ee
where $g_1(\omega_m)$ is the same function given in Eq.~(\ref{fwm}). As previously mentioned, Eq.~(\ref{bcs_tc_zz_app1})
can of course be written with a $-1/\lambda$ in the exponential, along with a prefactor denominator of $e^{3/2}$ instead of
$\sqrt{e}$. However, the present form more explicitly shows the role of the "normal-state" renormalization that gives rise to
the usual $1+\lambda$ factor, {\it along} with the not-so-usual $\sqrt{e}$ denominator in the pre-factor.

Written out explicitly, Eq.~(\ref{gapzz2}) reads
\bwt
\be
\Delta(\omega_m)  =  {1 \over 1 + {\bar{\omega}_m}^2} \left( 1 + \lambda \left[{3 \over 2}   - 
{1 \over 4 + \bar{\omega}_m^2}\left\{  
{2 - \bar{\omega}_m^2 \over \bar{\omega}_m}{\rm tan}^{-1}\bar{\omega}_m  - {3 \over 2} {\rm ln}(1 + \bar{\omega}_m^2) \right\}
- {1 \over |\bar{\omega}_m|}  
{\rm tan}^{-1}|\bar{\omega}_m| \right] \right).
\label{gapzz3}
\ee
\ewt
While Eqs.~(\ref{gapzz3}) looks very much like Eq.~(\ref{gap8}) with the $3/2$  vs. $1/2$ to account for the
$1+\lambda$ renormalization, there is one important difference: the large $\omega_m$ dependence for the first order term in
$\lambda$ is now $\approx (1/\omega_m^2)$ rather than $\approx (1/|\omega_m|)$ as was the case with $Z(\omega_m) = 1$.
Figures~4,~5,~and~6 essentially repeat the results of Figures~1,~2~and~3, respectively, now with $Z(\omega_m) \ne 1$.
Figure~4 shows already at these small values of $\lambda$ the detrimental effect of increased electron-phonon coupling that
arises through the normal scattering processes included in the normal part of the self-energy (included when $Z(\omega_m)$
is not equal to unity); this is apparent in the negative curvature of $T_c$ as a function of $\lambda$. In Fig.~5, where the gap
function is plotted as a function of Matsubara frequency, the results look qualitatively very similar to those in Fig.~2. Similarly,
in Fig.~6 the deviations from a decaying Lorentzian function look very similar to those in Fig.~3. The analytical results look
equally impressive, though in Fig.~6 the extra corrections from the renormalization function, $Z(\omega_m)$, are included, and the
decay at large frequency (not shown) is inversely as the square of the Matsubara frequency.

It is worth noting that with the explicit function of Matsubara frequency given by Eq.~(\ref{gapzz3}), an analytical continuation to
real frequency is straightforward. The Lorentzian on the imaginary axis now becomes a square root singularity on the real axis, with
the singularity occurring at the phonon frequency, once again highlighting that the gap function is definitely {\it not} constant for
frequencies up to the Einstein frequency, as in BCS theory. Additional gap structure as a function of frequency will arise in the term
proportional to $\lambda$, but this structure will of course be weak in this limit.

\section{Summary}

By now extensive solutions have been shown in innumerable papers for the gap function solution to the Eliashberg equations, as
indicated in the various reviews cited. In this paper we fill a hole in this tabulation, by presenting numerical solutions and analysis in
the weak coupling limit. The difficulty until now has been the number of Matsubara frequencies required for demonstrable
convergence. For example, we have used more than 120 000 (positive) Matsubara frequencies to achieve convergence for
some of the low electron-phonon couplings used in this study. We have also obtained analytical solutions to first order in
the coupling constant to reinforce these numerical solutions. The main messages of this study, reinforcing those of 
Refs.~[\onlinecite{karakozov76,wang13}] are\par
(i) the weak coupling expression for superconducting $T_c$ has a reduced pre-factor multiplying the phonon frequency scale,\par
(ii) the gap function approaches a Lorentzian function of frequency as $\lambda \rightarrow 0$, and first order corrections provide
very good, quantitatively correct results when compared to numerical results. This corrects the impression that the frequency
dependence of the order parameter is a feature that arises in Eliashberg theory {\it only beyond} the weak coupling regime. In
fact it remains a characteristic of the superconducting state even in the weak coupling limit, in contrast to the picture provided in
the BCS model calculation.\par
Further investigation will include results in the superconducting state, below $T_c$ and at zero temperature. In particular, the gap
edge at zero temperature, given in BCS theory by an analytical result similar to that of $T_c$ (Eq.~(\ref{bcs_tc} or \ref{bcs_tc_zz}),
will also acquire a correction in weak coupling Eliashberg theory analogous to that for $T_c$, i.e. 
Eq.~(\ref{bcs_tc_app1} or \ref{bcs_tc_zz_app1}), so that the gap ratio remains universal as $\lambda \rightarrow 0$.\cite{mitrovic84}
Another avenue of possible investigation, perhaps through the Josephson Effect, is to determine whether the frequency dependence
of the gap function can be measured, even in weakly coupled superconductors like Aluminium. 

\medskip

\noindent{\it Note added in proof:} We were alerted to $T_c$ solutions in the literature after this paper was submitted. In 
Ref.~[\onlinecite{combescot90}] expressions were derived for $T_c$ in the weak coupling limit for any shape of $\alpha^2F(\nu)$,
while in Ref.~[\onlinecite{freericks94}] the authors use a more general framework that nonetheless reproduces the correct prefactor
for $T_c$ in the weak coupling limit. We are grateful to Roland Combescot and Jim Freericks for bringing these papers to our attention.

\begin{acknowledgments}

This work was supported in part by the Natural Sciences and Engineering
Research Council of Canada (NSERC). We thank Andrey Chubukov, who first
brought this problem to our attention many years ago, and kindly provided us with introductory notes
for the derivation of Eq.~(\ref{bcs_tc_app1}).

\end{acknowledgments}

%


\begin{thebibliography}{99}

\bibitem{eliashberg60} G.M. Eliashberg, {\itt Interactions between Electrons and Lattice Vibrations in a Superconductor,} 
Zh. Eksperim. i Teor. Fiz. {\bf 38} 966 (1960); 
\href{http://www.w2agz.com/Library/Classic%20Papers%20in%20Superconductivity/Eliashberg,%20e-p%20Interactions%20in%20SCs,%20Sov-Phys%20JETP%2011,%20696%20(1960).pdf}
{Soviet Phys.  JETP {\bf 11} 696-702 (1960).}

\bibitem{bardeen57}
J. Bardeen, L.N. Cooper and J.R. Schrieffer, {\it Theory of Superconductivity},
\href{https://journals.aps.org/pr/abstract/10.1103/PhysRev.106.162}{Phys. Rev. {\bf 106}, 162 (1957)};
\href{https://journals.aps.org/pr/abstract/10.1103/PhysRev.108.1175}{Phys. Rev. {\bf 108}, 1175 (1957)}.

\bibitem{thouless60} D.J. Thouless, {\it Strong-Coupling Limit in the Theory of Superconductivity}, \href{https://journals.aps.org/pr/pdf/10.1103/PhysRev.117.1256}{Phys. Rev. {\bf 117}, 1256 (1960).}

\bibitem{swihart62} J.C. Swihart, {\it Solutions of the BCS Integral Equation and Deviations from
the Law of Corresponding States}, \href{https://ieeexplore-ieee-org/stamp/stamp.jsp?tp=&arnumber=5392405}{IBM Journal of
Research and Development {\bf 2}, 14 (1962).}

\bibitem{allen75} P.B. Allen and R.C. Dynes, {\it Transition temperature of strong-coupled superconductors reanalyzed},
\href{https://journals.aps.org/prb/abstract/10.1103/PhysRevB.12.905}{Phys. Rev. B{\bf 12} 905 (1975)}.

\bibitem{marsiglio91} F. Marsiglio and J.P. Carbotte, {\it Gap function and density of states in the 
strong-coupling limit for an electron-boson system},
\href{https://journals.aps.org/prb/abstract/10.1103/PhysRevB.43.5355}{Phys. Rev. B{\bf 43} 5355 (1991)}.  

\bibitem{remark_leggett} Zero temperature properties, at least for BCS theory is a different matter, as Leggett\cite{leggett80} first
emphasized. He pointed out that the BCS ground state is qualitatively correct even in the strong coupling limit. Extensions of
these ideas were presented in Ref.~[\onlinecite{nozieres85}] and have continued to the present.

\bibitem{leggett80} A.J. Leggett, {\it Cooper Pairing in Spin-polarized Fermi systems}, 
\href{https://jphyscol.journaldephysique.org/articles/jphyscol/abs/1980/07/jphyscol198041C704/jphyscol198041C704.html}
{J. de Physique, C7, {\bf 41}, 19 (1980)};\\
A.J. Leggett, {\it Diatomic Molecules and Cooper Pairs} in 
\href{https://link.springer.com/content/pdf/10.1007/BFb0120125.pdf}{{\it Modern Trends in the Theory of Condensed Matter}},
edited by S. Pekalski and J. Przystawa (Springer, Berlin, 1980)p. 13.

\bibitem{nozieres85}
P. Nozi\`eres and S. Schmitt-Rink, {\it Bose Condensation in an Attractive Fermion Gas:
From Weak to Strong Coupling Superconductivity}, 
\href{https://link.springer.com/article/10.1007/BF00683774}{J. Low Temp. Phys. {\bf 59}, 195 (1985)}.

\bibitem{mitrovic84} B. Mitrovi\'c, H.G. Zarate, and J.P. Carbotte, {\it The ratio $2\Delta_0/(k_B T_c$ within Eliashberg theory},
\href{https://journals.aps.org/prb/abstract/10.1103/PhysRevB.29.184}{Phys. Rev. B{\bf 29} 184 (1984)}.

\bibitem{marsiglio86} F. Marsiglio and J.P. Carbotte, {\it Strong-coupling corrections to Bardeen-Cooper-Schrieffer ratios},
\href{https://journals.aps.org/prb/abstract/10.1103/PhysRevB.33.6141}{Phys. Rev. B{\bf 33} 6141 (1986)}.

\bibitem{carbotte90} J.P. Carbotte, {\it Properties of boson-exchange superconductors},
\href{https://journals.aps.org/rmp/abstract/10.1103/RevModPhys.62.1027}{Rev. Mod. Phys. {\bf 62}, 1027-1157 (1990)}.

\bibitem{karakozov76} A.E. Karakozov, E.G. Maksimov and S.A. Mashkov, {\it Effect of the frequency dependence 
of the electron-phonon interaction spectral function on the thermodynamic
properties of superconductors}, [ZhETF 68, 1937 (1975)]
\href{http://www.jetp.ac.ru/cgi-bin/dn/e_041_05_0971.pdf}{Sov. Phys. JETP {\bf 41} 971 (1976)}. See also
O.V. Dolgov, I.I. Mazin, A.A. Golubov, S.Y. Savrasov, and E.G. Maksimov, {\it Critical Temperature and Enhanced 
Isotope Effect in the Presence of Paramagnons in Phonon-Mediated Superconductors},
\href{https://journals.aps.org/prl/abstract/10.1103/PhysRevLett.95.257003}{Phys. Rev. Lett. {\bf 95}, 257003 (2005)}.

\bibitem{wang13} Yuxuan Wang and Andrey Chubukov, {\it Quantum-critical pairing in electron-doped cuprates},
\href{https://journals.aps.org/prb/abstract/10.1103/PhysRevB.88.024516}{Phys. Rev. B{\bf 88}, 024516 (2013)}. 
See particularly Appendix A. We are grateful to Andrey Chubukov for sharing personal notes detailing this derivation
with us.

\bibitem{mcmillan68} 
W.L. McMillan, {\it Transition Temperature of Strong-Coupled Superconductors},
\href{https://journals.aps.org/pr/abstract/10.1103/PhysRev.167.331}{Phys. Rev. {\bf 167} 331 (1968)}

\bibitem{gorkov61} L.P. Gor'kov and T.K. Melik-Barkhudarov, {\it Contribution to the theory of superfluidity in an
imperfect Fermi gas}, J.~Exptl.~Theoret.~Phys. {\bf 40}, 1452 (1961);
\href{http://www.jetp.ac.ru/cgi-bin/e/index/e/13/5/p1018?a=list}{Sov. Phys. JETP {\bf 13}, 1018 (1961)}.

\bibitem{kohn65} W. Kohn and J.M. Luttinger, {\it New Mechanism for Superconductivity},
\href{https://journals.aps.org/prl/abstract/10.1103/PhysRevLett.15.524}{Phys. Rev. Lett. {\bf 15}, 524 (1965)};
J.M. Luttinger, {\it New Mechanism for Superconductivity},
\href{https://journals.aps.org/pr/abstract/10.1103/PhysRev.150.202}{Phys. Rev. {\bf 150}, 202 (1966)}.

\bibitem{wang13b} Andrey Chubukov, private communication. See also Ref.~[\onlinecite{wang13}] for an implementation
of the Kohn-Luttinger calculation into a mechanism proposed for the cuprates.

\bibitem{allen82} P.B. Allen and B. Mitrovi\'{c}, {\it Theory of Superconducting $T_c$},
in {\it Solid State Physics}, edited by H. Ehrenreich, F.~Seitz, and D. Turnbull (Academic, New York, 1982) Vol. 37, p.1.  

\bibitem{chubukov08} See, for example, 
A.V. Chubukov, D. Pines, and J. Schmalian, `A Spin Fluctuation Model for $d$-Wave Superconductivity',
Review Chapter in {\it Superconductivity, Conventional and Unconventional Superconductors},
edited by K.H. Bennemann and J.B. Ketterson (Springer-Verlag, Berlin, 2008), pp. 1349-1413, and
D. Manske, I. Eremin, and K.H. Bennemann, `Electronic Theory for Superconductivity in High-$T_c$ Cuprates and Sr$_2$RuO$_4$',
Review Chapter in {\it Superconductivity, Conventional and Unconventional Superconductors},
edited by K.H. Bennemann and J.B. Ketterson (Springer-Verlag, Berlin, 2008), pp. 1415-1515.

\bibitem{marsiglio08} F. Marsiglio and J.P. Carbotte, `Electron-Phonon Superconductivity',
Review Chapter in {\it Superconductivity, Conventional and Unconventional Superconductors},
edited by K.H. Bennemann and J.B. Ketterson (Springer-Verlag, Berlin, 2008), pp. 73-162.

\bibitem{marsiglio92} F. Marsiglio, \com{\itt Eliashberg Theory of the Critical Temperature and Isotope Effect. 
Dependence on Bandwidth, Band-Filling, and Direct Coulomb Repulsion,} 
\href{https://link.springer.com/article/10.1007/BF00118329]}{J. Low Temp. Phys. {\bf 87} 659-682 (1992)}.

\bibitem{remark_other} While the finite electronic bandwidth is irrelevant for superconducting $T_c$ in the weak coupling
limit, it does remain relevant for other, dynamic quantities even in this limit. See F. Do\u gan and F. Marsiglio,
{\it Self-consistent modification to the electron density of states due to electron-phonon coupling in metals}, 
\href{https://journals.aps.org/prb/abstract/10.1103/PhysRevB.68.165102}{Phys. Rev. {\bf B}68, 165102, (2003)}.  

\bibitem{owen71} C.S. Owen and D.J. Scalapino, {\it S-state instabilities for retarded interactions},
\href{https://doi.org/10.1016/0031-8914(71)90320-X}{Physica {\bf 55} 691 (1971)}.

\bibitem{bergmann73}
G. Bergmann and D. Rainer, {\it The Sensitivity of the Transition Temperature
to Changes in $\alpha^2F(\omega)$},
\href{https://link.springer.com/article/10.1007/BF02351862}{Z. Physik {\bf 263} 59 (1973)}.

\bibitem{rainer74}
D. Rainer and G. Bergmann, {\it Temperature-dependence of $H_{c2}$ and $\kappa_1$ in Strong Coupling Superconductors},
\href{https://link.springer.com/article/10.1007/BF00658876}{J. Low Temp. Phys. {\bf 14} 501 (1974)}.   

\bibitem{carbotte86} J.P. Carbotte, F. Marsiglio and B. Mitrovi\'c,
{\it Maximum $2\Delta_0/k_BT_c$ for electron-phonon superconductors},
\href{https://journals.aps.org/prb/abstract/10.1103/PhysRevB.33.6135}{Phys. Rev. B{\bf 33} 6135 (1986)}.

\bibitem{karakozov76b} See the un-numbered equation following Eq.~(29) in Ref.~(\onlinecite{karakozov76}), and in 
particular the first term of $1/2$ for $A$.

\bibitem{remark_prefactor} Speficially, instead of the usual pre-factor, $1.134 \equiv 2e^\gamma/\pi$, 
where $\gamma \equiv 0.5772...$ is Euler's constant, they obtained $1.134/e^{1/2}$.


\bibitem{cappelluti07} E. Cappelluti and G.A. Ummarino, {\it Strong-coupling properties of unbalanced Eliashberg superconductors},
\href{https://journals.aps.org/prb/abstract/10.1103/PhysRevB.76.104522}{Phys. Rev. B{\bf 76}, 104522 (2007)}.

\bibitem{abramowitz72} M. Abramowitz and I.A. Stegun, \href{http://people.math.sfu.ca/~cbm/aands/}
{\it Handbook of Mathematical Functions} (Dover, New York, 1972).

\bibitem{olver10}
Frank W.J. Olver, Daniel W. Lozier, Ronald F. Boisvert, and Charles W. Clark,
\href{https://www.researchgate.net/publication/234783338_NIST_Handbook_of_Mathematical_Functions}
{{\it NIST Handbook of Mathematical Functions}}, (Cambridge University Press, Cambridge, 2010).

\bibitem{remark_normalization} There are some very slight inconsistencies since our numerical solution is always
normalized so that $\Delta_{\rm num} (\omega_1) \equiv 1$, but these differences are not discernible on the scale of
the plots shown.

\bibitem{anderson59} P.W. Anderson, {\it Theory of Dirty Superconductors},
\href{https://doi.org/10.1016/0022-3697(59)90036-8}{J. Phys. Chem. Solids {\bf 11} 26 (1959)}.

\bibitem{combescot90} R. Combescot, {\it Critical temperature of superconductors: Exact solution from Eliashberg equations
on the weak-coupling side}, \href{https://journals.aps.org/prb/pdf/10.1103/PhysRevB.42.7810}{Phys. Rev. B{\bf 42}, 7810 (1990)}.

\bibitem{freericks94} J.K. Freericks and D.J. Scalapino, {\it Weak-coupling expansions for the attractive Holstein and Hubbard models},
\href{https://journals.aps.org/prb/pdf/10.1103/PhysRevB.49.6368}{Phys. Rev. B{\bf 49}, 6368 (1994)}.

\end{thebibliography}
\end{document}